\documentclass[a4paper,11pt]{article}%

\usepackage{amssymb,amsmath,amsfonts}
\usepackage{amsmath}
\usepackage{amsfonts}
\usepackage{amssymb}
\usepackage{graphicx}%
\providecommand{\U}[1]{\protect\rule{.1in}{.1in}}

\newtheorem{idea memo}[theorem]{Idea Memo}

\begin{document}

\title{New Interpretation of Equivalence Principle in General Relativity from the
viewpoint of Micro-Macro duality\thanks{Invited Talk at "Foundations of Probability and Physics 6"
(13--16 June 2011) at Linnaeus University, V\"{a}xj\"{o},
Sweden.}}

\author{Izumi Ojima\\
RIMS, Kyoto University, Kyoto 606-8502, Japan}
\date{}

\maketitle

\begin{abstract}
We propose a new interpretation of the equivalence principle underlying
Einstein's general relativity: a free-falling frame with gravitational force
eliminated locally in a small spacetime region shows the existence of a
boundary level, below which gravity is absent and above which \textit{gravity
emerges as condensation effect of microscopic motions} within each such frame
and interrelates free-falling frames at different spacetime points. In this
picture, gravitational field as a mediator of different free-falling frames
shows a remarkable \textit{parallelism with an order parameter to specify
\textquotedblleft degenerate vacua\textquotedblright} in different
thermodynamic pure phases
due to the \textit{condensation effects} in phase transitions.
As the physical basis of general relativity is found in the universality of
mass point motions due to the constancy of [inertial $m$]/[gravitational $m$],
the general relativistic notion of\ \textquotedblleft
spacetime\textquotedblright\ should be meaningful only in the validity regime
of this constancy, which is \textit{of empirical nature},\textit{\ }contrary
to the usual consensus. At the end, we comment on the impossibility to observe
gravitational waves which would make gravitons and quantum gravity questionable.

\end{abstract}

\qquad \qquad \quad Micro-Macro Duality

\noindent Classification: 04.20.Cv, 03.65.Ta, 03.70.+k, 02.10.-v

\section{Emergence of spacetime and gravity}

The motivating idea of Einstein's theory of general relativity is strongly of
geometric nature and naturally explains his enthusiasm about unsuccessful
attempts at \textquotedblleft Unified Field Theory\textquotedblright\ for
unifying electromagnetism and gravity. Except for some short period, the same
enthusiasm for \textquotedblleft Geometrization of Physics\textquotedblright%
\ has overwhelmingly dominated in particle physics for more than three
decades. From this viewpoint, however, the physical origin of gravity and
spacetime cannot be unveiled; what about the use of \textit{spacetime} notion
even in the situations with totally \textit{indeterminate future}??

To overcome such difficulties, we propose a scenario for deriving gravity and
spacetime as \textit{epigenetic} secondary notions \textit{emerging from
microscopic physics} of matter motions. For this purpose, the essence of this
report is just to explain the following diagram consisting of the structures
relevant to the emergence of special- and general-relativistic spacetimes (see below).

\vskip8pt The basic ingredients necessary for this purpose are as follows:

\vskip5pt \ i) {\textit{\textbf{Independence}}} = freely falling frames as
\textquotedblleft sectors\textquotedblright\ without gravity containing only
strong, weak \& electromagnetic couplings

\vskip5pt ii) {\textit{\textbf{Coupling}}}= gravitational force $\Gamma
_{\mu\nu}^{\lambda}$defined as Levi-Civita connection to connect different
free-falling fames as \textquotedblleft sectors\textquotedblright\ at the
\textit{meta-level, and},\textit{ }

\vskip5pt iii) {\textit{\textbf{Dependence}}}= \ the composite system arising
from the above physical systems constructed by three kinds of forces (strong,
weak \& electromagnetic) coupled with each other by the gravitational force.

\noindent
$%
\begin{array}
[c]{|c|c|c|}\hline%
\begin{array}
[c]{c}%
\text{\textit{\textbf{Spec = }}}\\
\text{spacetime}%
\end{array}
\text{ }\{x^{\mu}\} & \text{\ \ \ \ \textit{\textbf{Gen. Rel.}} }= &
\begin{array}
[c]{c}%
\text{General Cov. }\\
\text{Functorial Sym }\mathcal{G}%
\end{array}
\\\hline
g_{\mu\nu} & \Gamma_{\mu\nu}^{\lambda}\curvearrowright%
\begin{array}
[c]{c}%
\text{freely falling frames}\\
\text{at different }\{x^{\mu}\}
\end{array}
& \uparrow%
\begin{array}
[c]{c}%
\text{induced rep}\\
Ind_{H}^{\mathcal{G}}%
\end{array}
\\\hline%
\begin{array}
[c]{c}%
R,R_{\mu\nu}\\
\uparrow\downarrow
\end{array}
&
\begin{array}
[c]{c}%
m_{grav}=m_{inert}\text{ }\Uparrow\text{: \textit{\textbf{Equiv. principle}}%
}\\
\text{of inside \&~outside of a \textit{\textbf{sector}} }x^{\mu~}\\
\text{with~\textit{\textbf{no gravity}}}%
\end{array}
&
\begin{array}
[c]{c}%
\text{Unbroken Sym }\\
H=\mathcal{P}_{+}^{\uparrow}%
\end{array}
\\\hline%
\begin{array}
[c]{c}%
\text{\textit{\textbf{Einstein eqn}}}\\
R_{\mu\nu}-\dfrac{1}{2}g_{\mu\nu}R
\end{array}
& \left[
\begin{array}
[c]{c}%
x\\
\shortmid\\
p
\end{array}
~\text{duality}\right]  \text{ \ \ \ \ \ \ \ \ \ \ \ \ \ \ \ \ \ \ }%
\searrow\nwarrow &
\begin{array}
[c]{c}%
\text{local spacetime}\\
\text{\textit{\textbf{emergence}} \ }1/c
\end{array}
\\\hline%
\begin{array}
[c]{c}%
\text{\quad\quad}\parallel\\
\text{\quad\ \ \ \ \ \ }\kappa\omega(T_{\mu\nu})\\
\text{state}~\omega\text{: }\uparrow\downarrow
\end{array}
&
\begin{array}
[c]{c}%
\text{\ \ \ }(\text{material path})\\
\\
\text{\ W-S~angle\ \ }\theta_{WS}\swarrow\nearrow
\end{array}
&
\begin{array}
[c]{ccc}%
F_{\mu\nu} & \overset{\text{{\large Maxwell~eqn}}}{\leftarrow} & eJ_{\mu}\\
\uparrow\downarrow &  & \uparrow\downarrow\\
A_{\mu} & \underset{\text{\textit{\textbf{covar. der.}}}}{\rightarrow} & \psi
\end{array}
\\\hline
\text{\ \ \ \ \ \ \ \ \ \ \ }\uparrow\downarrow & \text{Weak
Interactions\ \ \ \ \ \ \ } & \\\hline
\text{ \ \ \ \ \ \ \ \ \ \ \ \ }T_{\mu\nu} & \text{: \textit{\textbf{Dynamics}%
}\ \ \ \ \ \ \ \ \ \ \ } & \text{Strong Interactions\ \ \ \ }\\\hline
\end{array}
$

\noindent By reviewing how general-relativistic spacetime emerges from the
physical processes in Micro quantum systems, we clarify here under which
condition the notion of \textquotedblleft spacetime\textquotedblright\ can be
meaningful from the viewpoint of \textquotedblleft Micro-Macro
duality\textquotedblright.

\vskip8pt \ {Micro-Macro Duality \& \textit{\textbf{Quadrality Scheme}}}

\vskip4pt \ 1)\ \textit{\textbf{Micro-Macro duality}} \cite{MicMac} as a
mathematical version of \textquotedblleft\textit{\textbf{quantum-classical
correpsondence}}\textquotedblright\ between microscopic
\textit{\textbf{sectors}} defined by quasi-equivalence (= unitary equiv. up to
multiplicity) classes of \textit{\textbf{factor states}} of observable algebra
\& macroscopic \textit{\textbf{inter-sectorial}} level described by
geometrical structures on the central spectrum $Spec(\mathfrak{Z})$:

\vskip6pt $%
\begin{array}
[c]{|cccc||c|}\hline
\longleftarrow%
\begin{array}
[c]{c}%
\text{Visible }\\
\text{\textit{\textbf{Macro}}~}%
\end{array}
\text{ of} &
\begin{array}
[c]{c}%
\text{\textit{\textbf{independent}}}\\
\text{\textit{\textbf{objects}}}%
\end{array}
\text{ } & \cdots & \longrightarrow &
\begin{array}
[c]{c}%
\text{\textit{\textbf{Inter-}}}\\
\text{\textit{\textbf{sectorial}}}%
\end{array}
\\\hline
\cdots\text{ \ \ \ }\gamma_{N}\text{ \ \ \quad} &
\text{\textit{\textbf{sectors~ }}}\gamma & \gamma_{2} & \gamma_{1} &
Spec(\mathfrak{Z})\\\hline\hline
\vdots & \vdots & \vdots & \vdots & \uparrow%
\begin{array}
[c]{c}%
\text{\textit{\textbf{Intra-}}}\\
\text{\textit{\textbf{sectorial}}}%
\end{array}
\\
\cdots\text{\quad}\pi_{\gamma_{N}}\text{\ \ \ \ \quad} & \pi_{\gamma} &
\pi_{\gamma_{2}}\text{\ \ \ } & \pi_{\gamma_{1}}\text{\ } & \parallel\\
\vdots & \vdots & \vdots & \vdots & \downarrow\text{\ \ }%
\begin{array}
[c]{c}%
\text{invisible}\\
\text{\textit{\textbf{Micro}}}%
\end{array}
\\\hline
\end{array}
$\noindent

\vskip6pt According to Fourier \& Galois dualities, these Micro \& Macro are
in duality, meaning that the data given at Macro level is derived from the
analysis on Micro and vice versa.

\vskip8pt \ 2)\ {Quadrality Scheme: }As a combination of two kinds of
Micro-Macro dualities in \textquotedblleft horizontal\textquotedblright\ and
\textquotedblleft vertical\textquotedblright\ directions, a general
methodological framework can be formulated for theoretical description of
physical phenomena which I call \textit{\textbf{quadrality scheme}}:

\vskip8pt \noindent\noindent$%
\begin{array}
[c]{|c|c|c|c|c|}\hline
\text{Micro: }%
\begin{array}
[c]{c}%
\text{visible}\\
\text{levels}%
\end{array}
&  &
\begin{array}
[c]{c}%
\text{\textit{\textbf{Spec }}= }\\
\text{classifying space}%
\end{array}
&  & \\\hline%
\begin{array}
[c]{c}%
\text{classification}\\
\text{/emergence}%
\end{array}
& \nearrow & \text{{\small dual}}\uparrow\downarrow &  & \\\hline
\text{\textit{\textbf{States~\textit{\&~}Rep.'s}}} & \overset{\text{dual}%
}{\leftrightarrows} &
\begin{array}
[c]{c}%
\text{Fourier-Galois}\\
\text{dualities}%
\end{array}
& \overset{\text{dual}}{\leftrightarrows} &
\begin{array}
[c]{c}%
\text{\textit{\textbf{Alg}}ebra of}\\
\text{observables}%
\end{array}
\\\hline
&  & \text{{\small dual}}\uparrow\downarrow & \nearrow & \\\hline
&  & \text{ \ \ \ \ \ \ \ \ \ \ \textit{\textbf{Dyn}}amics} &  &
\begin{array}
[c]{c}%
\text{object }\\
\text{system}%
\end{array}
\text{: Micro}\\\hline
\end{array}
$

\vskip8pt In a sense, the essence of this scheme overlaps with the basic
structure controlling the four interactions as follows: \vskip8pt $%
\begin{array}
[c]{|c|c|c|}\hline
& \text{Gravity} & \\\hline
\text{Electromagnetism} & \text{(}%
\begin{array}
[c]{c}%
\text{Thermality }k_{B}\\
\leftrightarrows\text{\textquotedblleft Quantum\textquotedblright~}\hbar
\end{array}
\text{)} & \text{Weak forces}\\\hline
& \text{Strong force} & \\\hline
\end{array}
.$ \vskip6pt Namely, the meaning of \textquotedblleft unification of four
forces\textquotedblright\ need not be restricted to a simple-minded version
like the currently prevailing one with their convergence into a single entity,
but, the mutual relations among them may well alternatively be understood in
such a form of their integrated organization that they occupy mutually
different places in nature and in theoretical frameworks, playing different
roles inherent in each, through which the unified totality of nature and its
theoretical explanations can be attained. The standard picture of
\textquotedblleft unification\textquotedblright\ pursued in the context of
\textquotedblleft geometrization of physics\textquotedblright\ seems to lack
systematically considerations into this kind of aspects.

\vskip6pt

3) Fourier-Galois Duality: {Bi-directionality between Induction \& Deduction}

Essence of duality in Fourier transform $(\mathcal{F}f)(\gamma)=\int
\overline{\gamma(g)}f(g)dg)$\ ( $f\in L^{1}(G)$) is formulated by
Fourier-Pontryagin duality $G\rightleftarrows\hat{G}$ between a locally
compact abelian group $G$ and its dual group $\hat{G}$ consisting of all the
characters $\chi:G\rightarrow\mathbb{T}$. Via extension to compact cases due
to Tannaka \& Krein, the most general form can now be found in
Tatsuuma-Enock-Schwartz theorem of the duality between a locally compact
non-abelian group $G$ and its representation category $Rep(G)$ consisting of
\textquotedblleft all\textquotedblright\ the representations. The
corresponding version is formulated by Takesaki (and/or Takai) for dynamical
system with a (non-commutative) algebra $\mathcal{F}$ and with an action
$\tau:\mathcal{F\curvearrowleft}G$ of $G$ on $\mathcal{F}$ in such a form (in
C*- or W*-versions, respectively) as
\begin{align*}
\mathcal{F}^{G}\rtimes_{\hat{\tau}}\hat{G}  &  =\mathcal{F}\text{ \ \ \ \ :
{\textit{\textbf{Recovery of}}}\ }\mathcal{F}\text{\ {\textit{\textbf{from}}}
}G\text{-{\textit{\textbf{invariants}}} }\mathcal{F}^{G}\text{;}\\
\mathcal{F}\rtimes_{\tau}G  &  =\mathcal{F}^{G}\otimes\mathcal{K}%
(L^{2}(G))\text{ or }\mathcal{F}^{G}\otimes B(L^{2}(G)).
\end{align*}

In all cases, the Kac-Takesaki operators play crucial roles, in term of which
a method for {constructing composite system can be developed }systematically
{via coupling between object system and reference system. Unfortuately, we
should omit them here for lack of space. }

\vskip8pt \ 4)\ {Symmetry Breaking, condensed states} \& induced
representations $\rightarrow$ condensation and degenerate vacua

\vskip5pt Breakdown of a symmetry $G$ of a dynamical system $\mathcal{F}%
\curvearrowleft G$ in a state $\omega\in E_{\mathcal{F}}$ is characterized
\cite{Unif03} by non-invariance of the \textquotedblleft central
extension\textquotedblright\ of $\omega$ on the centre $\mathfrak{Z}%
_{\pi_{\omega}}(\mathcal{F}):=\pi_{\omega}(\mathcal{F})^{\prime\prime}\cap
\pi_{\omega}(\mathcal{F})^{\prime}$ under the corresponding $G$-action on
$\mathfrak{Z}_{\pi_{\omega}}(\mathcal{F})$. In this case, Galois closedness of
$\mathcal{F}^{G}$ is broken, which is recovered by dynamical system
$\mathcal{F}\curvearrowleft H$ described by a compact Lie subgroup $H$ of $G$
corresponding to unbroken symmetry: $\mathcal{F}=\mathcal{F}^{H}%
\rtimes\widehat{H}$. Then, the sector structure is determined by factor
spectrum $\overset{\frown}{\mathcal{F}^{H}}=Spec(\mathfrak{Z}(\mathcal{F}%
^{H}))=\hat{H}$: group dual consisting of irreducible unitary rep.'s of $H$.

\vskip8pt 5) Emergence of Macro by \textit{\textbf{condensation effects }}of
Micro $\rightarrow$ physical application of \textit{\textbf{forcing method }}
(which implies Born rule \cite{IO10})

\vskip5pt With $\widetilde{\mathcal{F}}:=\mathcal{F}^{H}\rtimes\widehat
{G}=\mathcal{F}\rtimes\widehat{(H\backslash G)}$ called an
\textit{\textbf{augmented algebra}} \cite{Unif03}, we have a
\textit{\textbf{split }}bundle exact sequence $\mathcal{F}^{H}\overset
{\tilde{m}}{\underset{\mathcal{\hookrightarrow}}{\twoheadleftarrow}}%
\widetilde{\mathcal{F}}\underset{\twoheadrightarrow}{\hookleftarrow}%
\widetilde{\mathcal{F}}/\mathcal{F}^{H}\simeq\widehat{G}$.

In this situation, \textit{\textbf{minimality}} of\textit{\textbf{\ }}$G$ and
$\widetilde{\mathcal{F}}$\ is guaranteed by $G$-\textit{\textbf{central
ergodicity}}, i.e., $G$-ergodicity of centre $\mathfrak{Z}_{\tilde{\pi}%
}(\widetilde{\mathcal{F}})$ in the rep. $\tilde{\pi}$ given by GNS rep. of
$\omega_{0}\circ\tilde{m}$ induced from vac. state $\omega_{0}$ of
$\mathcal{F}^{H}$ \cite{Unif03}, and we have the following commutativity
diagrams: \vskip7pt $%
\begin{array}
[c]{|c|c|c|}\hline
& ^{\text{1:1}}\swarrow\mathcal{F}^{H}=\widetilde{\mathcal{F}}^{G}\text{:
unbroken obs.} & \searrow^{\text{1:1}}\text{ \ \ \ \ \ }\\\hline
\mathcal{F} & \searrow\searrow^{\text{1:1}}\text{\ \ \ \ \ \ }\Downarrow
^{\text{1:1}}\text{ \ \ \ \ \ \ \ \ \ \ \ }^{\text{1:1}}\swarrow &
\widetilde{\mathcal{F}}^{H}\text{: }%
\begin{array}
[c]{c}%
\text{extended}\\
\text{obs. alg.}%
\end{array}
\\\hline
^{\text{onto}}\downarrow & \text{ \ \ \ \ \ \ \ \ \ \ \ \ \ \ \ }%
\widetilde{\mathcal{F}}\text{: augmented alg.\ } & \text{\ }\downarrow
^{\text{onto}}\\\hline
^{\text{onto}}\downarrow & \swarrow^{\text{onto}}\text{\ \ \ \ \ }%
\Downarrow^{\text{onto}}\text{ \ \ \ \ }^{\text{onto}}\searrow\searrow &
\text{\ }\downarrow^{\text{onto}}\\\hline
\text{ \ \ \ \ }\widehat{H} & \twoheadleftarrow\text{ \ \ \ \ \ }\widehat
{G}\text{ \ \ \ \ \ \ \ \ \ \ \ \ \ }\hookleftarrow & \text{\ }\widehat
{G/H}\text{ \ \ \ \ \ }\\\hline
\end{array}
$ \vskip7pt The above diagram for algebra extension is dual to the following
one for sectors: \vskip6pt $%
\begin{array}
[c]{|c|c|c|}\hline
& \overset{\frown}{\widetilde{\mathcal{F}}^{G}}=\text{ }\overset{\frown
}{\mathcal{F}^{H}}\simeq\text{\ }\widehat{H} & \text{: unbroken\ sectors}%
\\\hline
& \nearrow^{\text{onto}}\text{\ \ \ }\Uparrow\text{\ \ \ \ \ \ \ \ }%
\nwarrow^{\text{onto}}\text{ \ \ \ \ \ } & \Downarrow^{\text{1:1}}\\\hline
\overset{\frown}{\mathcal{F}} & \text{\ \ \ \ \ \ \ \ \ \ \ \ }\Uparrow
^{\text{onto}}\text{\ \ \ }\overset{\frown}{\widetilde{\mathcal{F}}^{H}}\simeq
G\underset{H}{\times}\widehat{H} & \text{: sector bdle \ \ \ \ \ \ }\\\hline
^{\text{1:1}}\uparrow & \nwarrow\nwarrow^{\text{onto}}\Uparrow
\text{\ \ \ \ \ \ \ }\nearrow^{\text{onto}}\text{\ \ \ \ }\uparrow &
\Downarrow\text{ \ \ \ }\\\hline
^{\text{1:1}}\uparrow & \text{\ \ \ \ \ \ \ \ \ \ \ }\overset{\frown
}{\widetilde{\mathcal{F}}}\text{ \ \ \ \ \ \ \ \ \ \ \ \ \ \ \ \ \ }%
^{\text{1:1}}\uparrow & \Downarrow^{\text{onto}}\\\hline
^{\text{1:1}}\uparrow & \nearrow^{\text{1:1}}\text{\ \ \ \ }\Uparrow
^{\text{1:1}}\text{\ \ \ \ }\nwarrow\nwarrow^{\text{1:1}}\text{\ \ }\uparrow &
\Downarrow\text{ \ \ \ }\\\hline
H & \hookrightarrow\text{\ \ \ \ \ \ }G\text{: broken\ \ \ \ }%
\twoheadrightarrow\text{ \ \ }G/H & \text{: deg. vacua\ \ \ \ \ \ \ }\\\hline
\end{array}
,$\newline where $\overset{\frown}{\mathcal{F}}=Spec(\mathfrak{Z}%
(\mathcal{F}))$ denotes the factor spectrum of $\mathcal{F}$, etc. \vskip9pt
{\textquotedblleft Sector Bundle'\ associated with Broken Symmetry}

\vskip6pt The physical essence of extension $\mathcal{F}^{G}\Longrightarrow
\mathcal{F}^{H}$ from the $G$-fixed point subalgebra $\mathcal{F}^{G}$ to
$H$-fixed one $\mathcal{F}^{H}$ can now be interpreted as \textquotedblleft
extension of coefficient algebra\ $\mathcal{F}^{G}$\textquotedblright\ by (the
dual of) $G/H$ to parametrize degenerate vacua: $\mathcal{F}^{H}%
=\widetilde{\mathcal{F}}^{G}=[(\mathcal{F}\rtimes\widehat{(H\backslash
G)}]^{G}=\mathcal{F}^{G}\rtimes\widehat{(H\backslash G)}$.

In this extension, a part $G/H$~of originally \textit{\textbf{invisible }}$G$
has become \textit{\textbf{visible }}through the \textit{\textbf{emergence of
degenerate vacua}} parametrized by\textit{\textbf{\ }}$G/H$\textit{\textbf{\ }%
}due to \textit{\textbf{condensation of order parameter }}$\in G/H $
associated with \textbf{S}(ponteneous) \textbf{S}(ymmetry) \textbf{B}(reaking)
of $G$ to $H$.

As a result, observables $A\in\mathcal{A}$ acquire $G/H$-dependence:
$\widetilde{A}=(G/H\ni\dot{g}\longmapsto\widetilde{A}(\dot{g})\in
\mathcal{A})\in\mathcal{A}\rtimes\widehat{(H\backslash G)}$, which should just
be interpreted as an example of \textit{\textbf{logical extension }%
}\cite{OjiOza} transforming a \textquotedblleft\textit{\textbf{constant}}
object\textquotedblright\ ($A\in\mathcal{A}$) into a \textquotedblleft%
\textit{\textbf{variable}} object\textquotedblright\ ($\widetilde{A}%
\in\mathcal{A}\rtimes\widehat{(H\backslash G)}$) having
\textbf{\textit{functional dependence}} on the
universal\textbf{\textit{\ classifying space}} $G/H$\ for multi-valued
\textbf{\textit{semantics}}(, as is familiar in non-standard and
Boolean-valued analysis).

\vskip8pt \ 6) {Emergence of Spacetime as Symmetry Breaking}

By \textbf{\textit{replacing}} $G/H$ \textbf{\textit{with spacetime}}, the
above situation can be regarded as a prototype for the origin of functional
dependence of physical quantities on spacetime coordinates, due to the
physical emergence of spacetime from microscopic physical world.

Along this line, we prescribe the similar logical extension procedure on the
observable algebra $\mathcal{F}^{H}$ adding $G/H$-dependence: $\mathcal{F}%
^{H}\rtimes\widehat{(H\backslash G)}=(\mathcal{F}\rtimes\widehat{(H\backslash
G)})^{H}=\widetilde{\mathcal{F}}^{H}.$ The whole sector structure of
$\widetilde{\mathcal{F}}^{H}=(\mathcal{F}^{H}\rtimes\widehat{(H\backslash
G)})$ can be identified with its factor spectrum $\overset{\frown}%
{\widetilde{\mathcal{F}}^{H}}=G\underset{H}{\times}\hat{H}$; this constitutes
a \textit{\textbf{sector bundle}}, $\hat{H}\hookrightarrow\overset{\frown
}{\widetilde{\mathcal{F}}^{H}}=G\underset{H}{\times}\hat{H}\twoheadrightarrow
G/H$, consisting of the classifying space $G/H$ of \textit{\textbf{degenerate
vacua}}, each fibre over which describes the sector structure $\hat{H}$ of
\textit{\textbf{unbroken}} remaining symmetry $H$ (or, more precisely, the
conjugated group $gHg^{-1}$ for the vacuum parametrized by $\dot{g}=gH\in G/H$).

Namely, \textit{\textbf{sector bundle}}, $\hat{H}\hookrightarrow
\overset{\frown}{\widetilde{\mathcal{F}}^{H}}=G\underset{H}{\times}\hat
{H}\twoheadrightarrow G/H$, can be seen as the \textit{\textbf{connection}}=
\textit{\textbf{splitting}} of bundle exact sequence dual to $\overset{\frown
}{\mathcal{F}^{H}}=\hat{H}\twoheadleftarrow\overset{\frown}{\widetilde
{\mathcal{F}}^{H}}=G\underset{H}{\times}\hat{H}\hookleftarrow G/H$ of
observable triples, $\mathcal{F}^{H}\hookrightarrow\widetilde{\mathcal{F}}%
^{H}=\mathcal{F}^{H}\rtimes\widehat{(H\backslash G)}\twoheadrightarrow
\widehat{(H\backslash G)}$!

Now we apply the above scheme to the situation with group $G$ containing both
\textit{external} (= spacetime) and \textit{internal} symmetries. \newline For
simplicity, the latter component described by a subgroup $H$ of $G$ is assumed
to be unbroken, and hence, the broken symmetry described by $G/H$ represents
the spacetime structure. It would be convenient to take $H$ as a normal
subgroup of $G$, while not essential. To be precise, $G/H$ may contain such
non-commutative components as spatial rotations (and Lorentz boosts) acting on
spacetime, but, we simply neglect this aspect to identify $G/H$ as spacetime
itself (from which the corresponding transformation group can easily be
recovered). \newline

Then, by identifying $G/H$ with a spacetime domain $\mathcal{R}$, we find an
impressive parallelism between the commutative diagram in the previous section
and the diagram in Doplicher-Roberts reconstruction \cite{DR90} of local field
net $\mathcal{R}\longmapsto\mathcal{F}(\mathcal{R})$ from local observable net
$\mathcal{R}\longmapsto\mathcal{A}(\mathcal{R})$ (without the two bottom
lines) as follows: \vskip6pt $%
\begin{array}
[c]{c}%
_{H}\swarrow\mathcal{F}^{H}=\widetilde{\mathcal{F}}^{G}\searrow_{G/H}\\
\mathcal{F}\text{\ \ \ \ \ \ \ \ \ \ \ \ \ }\Downarrow
\text{\ \ \ \ \ \ \ \ \ \ \ \ }\widetilde{\mathcal{F}}^{H}\\
\downarrow\text{\ }_{G/H}\searrow\searrow\text{\ \ }\widetilde{\mathcal{F}%
}\text{\ \ \ \ \ \ \ }\swarrow_{H}\downarrow\\
\downarrow\text{\ \ \ \ }\swarrow\text{\ \ \ \ }\Downarrow\text{\ \ \ }%
\searrow\searrow\text{\ \ \ }\downarrow\\
\widehat{H}\text{\ \ }\twoheadleftarrow\text{\ \ \ \ \ \ }\widehat{G}\text{
\ \ \ }\hookleftarrow\text{ \ \ }\widehat{G/H}%
\end{array}
\rightleftarrows%
\begin{array}
[c]{c}%
_{G}\swarrow\mathcal{O}_{\rho}=\mathcal{O}_{d}^{G}\searrow_{\mathcal{R}}\\
\mathcal{O}_{d}\text{\ \ \ \ \ \ \ \ \ \ }\Downarrow\text{\ \ \ \ \ \ \ \ }%
\mathcal{A}(\mathcal{R})\\
\downarrow\text{ }_{\mathcal{R}}\searrow\searrow\text{ \ }\mathcal{F}%
(\mathcal{R})\text{\ \ }\swarrow_{G}\text{ \ }\downarrow\\
\downarrow\text{\ \ \ \ }\swarrow\text{\ \ \ \ }\Downarrow\text{\ \ \ \ }%
\searrow\searrow\text{\ }\downarrow\\
\widehat{G}\text{\ \ }\twoheadleftarrow\text{\ \ \ \ }\widehat{G\times
\mathcal{R}}\text{ \ \ }\hookleftarrow\text{ \ \ \ }\widehat{\mathcal{R}}%
\end{array}
$, \vskip6pt where $\mathcal{O}_{d}$ is a Cuntz algebra of $d$-isometries.

Thus we have arrived at the stage just before gravity to be switched on, to
enter General Relativity via \textit{\textbf{Equivalence Principle}}. This can naturally
be formulated and understood by the above scheme in combination with induced
representation. So, we should recall here the diagram at the beginning.

\section{Physical meaning of Equivalence Principle in General Relativity in
the emergence process}

We consider processes of spacetime emergence taking place in parallel under
the influence of strong and electro-weak interactions other than gravity, each
of which results in a \textquotedblleft fiber\textquotedblright\ (= sector=
pure phase) parametrized by spacetime coordinates $x^{\mu}$. The word
\textquotedblleft fiber\textquotedblright\ here means a \textit{\textbf{flat
tangent space}} as a fiber ($T_{x}(M)$) of a tangent bundle ($T(M)$) on each
point \ $x^{\mu}$\ of the base space as a \textquotedblleft spacetime
manifold\textquotedblright\ (which \textit{would} be called $M$ but which
cannot be recognized yet as such); as its physics is controlled by the three
interactions other than the gravity, this fiber describes a
\textit{\textbf{free-falling frame }}without any gravitational force (the last
of which has not emerged yet). In connection with our discussion up to this
point, the word \textquotedblleft sector\textquotedblright\ should be more
appropriate than \textquotedblleft fiber\textquotedblright, we accept the
latter use in order to avoid the misunderstanding of what we are concerned
with here. To be precise, what we know up to now is only the simultaneous
processes of spacetime emergence at many \textquotedblleft
fiber\textquotedblright\ points $x^{\mu}$ each of which consists of the
physical world of Poincar\'{e} covariant quantum fields\ governed by the
strong and electro-weak interactions in the Minkowski spacetime but we do not
know anything about the mutual relations among different fibers. By picking up
just one specific \textquotedblleft fiber\textquotedblright, we focus on the
local physics described by the Poincar\'{e} covariant QFT developed inside of
the \textquotedblleft gravitation-less free falling system\textquotedblright,
which is nothing but the physical contents of \textquotedblleft%
\textit{\textbf{tangential world}}\textquotedblright\ equipped with local
Lorentz structure, on (or \textquotedblleft in\textquotedblright?) a\ point
$(x^{\mu})$ in the emergent \textquotedblleft base space\textquotedblright.

\vskip12pt

Now, we pose a question: what does it mean to impose the physical requirement
of \textquotedblleft equivalence principle\textquotedblright\ between
gravitational and inertial masses, $m_{grav}=m_{inert}$ on the situation after
the \textquotedblleft individual\textquotedblright\ processes of free-falling
systems arising from the emergence of special-relativistic local spacetime?
While the notion of \textquotedblleft inertial mass\textquotedblright\ already
exists in the \textquotedblleft standard\textquotedblright\ physics formulated
within the free-falling frames without gravity, it does not apply to the case
of \textquotedblleft gravitational mass\textquotedblright\ before our starting
to discuss the situations governed by the gravitational interaction. It can be
meaningful only in the context where such an attribute is assigned to a(n
asymptotically) free mass point on the mass-shell, as generating the
gravitational force or field as the forth one other than strong and
electro-weak forces, through Einstein's gravitational equation:
\[
R_{\mu\nu}-\dfrac{1}{2}g_{\mu\nu}R=\kappa T_{\mu\nu}.
\]
When we find the first (or, the 0-th approximated) roles of gravity in
regulating the mutual relations among different fibers= sectors as
free-falling frames, the proper range of action of the gravitational mass
$m_{grav}$ is at the level of \textquotedblleft inter-fiber= inter-sectorial
relations\textquotedblright, but, in contrast, that of the inertial one
$m_{inert}$ is in the physics within each \textquotedblleft
fiber\textquotedblright\ (or sector). Therefore, the equivalence principle
qualitatively controls in a bi-directional way the mutual duality relation
between the inside and the outside of \textquotedblleft
fibers\textquotedblright\ (or, sectors)\footnote{From the viewpoint of
emergence as a process of phase separation, the roles played by the
free-falling frame in each \textquotedblleft fiber\textquotedblright\ and by
\textquotedblleft base space\textquotedblright\ can be compared with $H$ and
$G/H$ whose duality relation can be seen in the form of \textquotedblleft
Helgason duality\textquotedblright. In this sense, the gravitational
equivalence principle is analogous to \textquotedblleft Helgason
duality\textquotedblright.}: we suppose that the inter-fiber relation of
free-falling frames on the \textquotedblleft neighbouring\textquotedblright%
\ points $x^{\mu}$ and $x^{\mu}+\delta x^{\mu}$ is controlled by the
connection coefficients$\ \Gamma_{\mu\nu}^{\lambda}$, as is indicated in the
diagram at the beginning, which results in a force propertional to the
gravitational mass $m_{grav}$~acting on the inertial mass $m_{inert}$. Then,
the Newtonian equation of motion of the mass point $m_{inert}$ with the
velocity vector $v^{\lambda}:=\dfrac{dx^{\lambda}}{d\tau}$ can be written as,
\[
m_{inert}dv^{\lambda}=-v^{\mu}(m_{grav}\Gamma_{\mu\nu}^{\lambda}dx^{\nu
})=m_{grav}v^{\mu}\nabla_{\mu}dx^{\lambda}.
\]
By the requirement of equivalence principle\textit{\textbf{ }}$m_{grav}%
=m_{inert}$, this reduces to the \textit{\textbf{geodesic equation}},
$\dfrac{dv^{\lambda}}{d\tau}+\Gamma_{\mu\nu}^{\lambda}v^{\mu}v^{\nu}=0,$whose
\textit{\textbf{purely geometric}} form and independence of the specific mass
values ensure the universality of the mass-point motions. Namely, through the
validity of \textit{\textbf{equivalence principle }}$m_{grav}=m_{inert}$, the
spacetime notion $x^{\mu}$~acquires its own abstract universal meaning,
independently of its physical origin in the mutual relations among different
\textquotedblleft fibers\textquotedblright\ of local physics consisting of
three interactions, to such an extreme extent that space and \textquotedblleft
time\textquotedblright\ exist in themselves, extending from the past, the
present and even the future! Eventually, the physical motions of mass points
are now absorbed into a (small) part of spacetime geometry in the form of
geodesic motions, without exhibiting their individuality. Owing to this
mechanism, we can easily forget about the \textit{\textbf{physical origin}} of
spacetime, which can, however, exhibit its existence in the situation where
the validity of equivalence principle is threatened. It is also interesting to
note that the above equation of motion can be rewritten in terms of momentum
$p^{\lambda}=mv^{\lambda}$ into
\[
dp^{\lambda}=p^{\mu}\nabla_{\mu}dx^{\lambda},
\]
which explains that mass-point motions as geodesic motion can be absorbed into
the covariance (of physical motions) under the (covariantized) general
coordinate transformations. If the above discussion is compared with the
standard mathematical treatment of bundle structures in differential geometry,
we understand that ours go from physics in the (standard) fiber to the
mathematical structure of tbe bundle and base spaces in the opposite direction
to the latter and that the mathematical essence of the equivalence principle
lies in the $G$-structure of the tangent and frame bundles of the spacetime
$M$ with $G$ being identified with the Lorentz group \cite{Dieu}.

It would also be of interest to compare the above situation of gravity with
that of electromagnetism: in this case, once the (field strength of)
electromagnetic field $F_{\mu\nu}$ as a universal quantity is generated,
$F_{\mu\nu}\leftarrow J_{\mu}$, via the Maxwell equation, $\partial^{\nu
}F_{\mu\nu}=eJ_{\mu},$ from the electric current $J_{\mu}$ arising from the
microscopic matter motions, the resulting $F_{\mu\nu}$ starts to control all
the matter motions with a 4-velocity $v^{\mu}$\ by the Lorentz force
$eF_{\mu\nu}v^{\nu}$ acting on them, through which the coupled system of
electromagnetic field and matter motions is equationally closed. In the
direction from matter motions to universal quantities, $R_{\mu\nu}$ \&
$g_{\mu\nu}$ $\leftarrow T_{\mu\nu}$ [: to Macro, or meta-level from Micro],
the case of gravity shares the common features with the above
electromagnetism, through a well-known form of the Einstein equation,
$R_{\mu\nu}-\dfrac{1}{2}g_{\mu\nu}R=\kappa T_{\mu\nu},$ where geometric
quantities, $R_{\mu\nu}$ \& $g_{\mu\nu}$, related with gravity is generated
from the energy-momentum tensor $T_{\mu\nu}$ of matter motions. In the
opposite direction of the gravitational field $R_{\mu\nu}$ \& $g_{\mu\nu}$ to
exert its counter-action on matter motions as its sources, however, our
considerations above exhibit certain complicated aspects involved, in such
forms as the essential roles played by the formation of spacetime points
$x^{\mu}$ and the action of $\Gamma_{\mu\nu}^{\lambda}$, whose essence cannot
be exhausted in the direction from Macro to Micro levels. In other words, what
determines how the generated gravity counter-acts on matter motions is not
only the spacetime point $x^{\mu}$ emerging as the indices of a family of
free-falling systems, but also the equivalence principle, $m_{grav}=m_{inert}%
$, to equate the gravitational mass $m_{grav}$ appearing as the \textit{source
of gravitation field} in the Einstein equation and the \textit{inertial mass}
$m_{inert}$ characterizing the Newtonian-mechanical mass point summarizing the
(micro-)physical contents in the free-falling system indexed by $x^{\mu}$. By
this equivalence principle, the latter \textit{physical} meaning related with
Newtonian mechanics within each fiber (: Micro) is \textit{absorbed} into the
purely \textit{geometric} context of geodesics on Macro level, which finally
settles the physical and geometric meanings of general-relativistic
\textquotedblleft spacetime\textquotedblright\ and \textquotedblleft
gravity\textquotedblright.

The final pictures attained~in both cases, however, are quite similar, in such
forms as
\[
dp_{\mu}=p^{\nu}\dfrac{e}{m}(F_{\nu\mu}d\tau);\text{ \ \ \ \ }dp^{\lambda
}=p^{\mu}\text{ }(\nabla_{\mu}dx^{\lambda})\text{\ \ \ (if }m_{grav}%
=m_{inert}\text{),}%
\]
in the cases of electromagnetic Lorentz force and gravitational force,
respectively. Their common features can be seen in the parallelism with the
left action of a Lie group $G$ on the homogeneous space $G/H$ by the left
$G$-shift: $G/H\ni sH\longmapsto g(sH)=gsH\in G/H$, which can physically be
interpreted as the action of a broken symmetry $G$ with its subgroup $H$
remaining unbroken taken care of by the Goldstone modes $\thicksim G/H$ acting
on the space $G/H$ of degenerate vacua arising from the condensation effect
due to the symmetry breaking. Namely, both the electromagnetic force
$F_{\nu\mu}$ and the gravitational force $\Gamma_{\mu\nu}^{\lambda}$ behave as
Goldstone-like modes acting transitively on the emergent classical Macro
objects arising from the condensation effects due to some symmetry breakings.
While the question may be subtle as to what kind of symmetry is broken in
electromagnetism, something related to local gauge invariance is broken,
triggering the emergence of the Minkowski spacetime as the condensation, whose
Goldstone mode is given by $F_{\mu\nu}$ (or, $eF_{\mu\nu}/m$). In the case of
gravity, what is broken is the invariance under the general coordinate
transformations, the emerging condensation is the spacetime coordinates
$x^{\mu}$ to parametrize the free-falling frames, with Goldstone mode being
the Levi-Civita connection $\Gamma_{\mu\nu}^{\lambda}$ (or, $\nabla_{\mu
}dx^{\lambda}=-\Gamma_{\mu\nu}^{\lambda}dx^{\nu}$).

\section{Absence of gravitational sink and of gravitational waves}
At the end, we add brief comments on a new observation about the
absence of gravitational waves, which can be seen as follows:

1) In the general theory of relativity and in all modern physics, the notion
of spacetime point $x^{\mu}$ occupies the \textit{most fundamental }position
to parametrize all the events taking place in macroscopic nature which can be
compared with the \textquotedblleft\textit{elements}\textquotedblright\ in set
theory, upon which (almost) all the structures are built. Combining this
aspect with the processes of emergence from Micro to Macro, spacetime points
$x^{\mu}$ can be interpreted as \textquotedblleft\textit{initial
objects}\textquotedblright\textit{ }in category theory\footnote{The importance
of the notions of intial and terminal objects has been emphasized by Dr. H.
Saigo in combination with that of the category of sets, to whom I am very
grateful for useful discussions.} characterized by the uniqueness of arrow
emanating from it, with all other arrows convergent to it: in this context,
there is such a sharp asymmetry between the two sides of a spacetime point
$x^{\mu}$, that there are many arrows \textit{\textbf{to}} $x^{\mu}$ involving
three interactions \textit{\textbf{other than gravity}} but that arrows
involving gravity are only those unique ones emanating \textit{\textbf{from }%
}each $x^{\mu}$.~

2) The above asymmetry implies the \textit{\textbf{absence of gravitational
sinks }}to which many arrows involving gravity would converge. Without this
asymmetry, the qualification of $x^{\mu}$ as set-theoretical elements could
not be consistent with the regularity axiom in set theory denying the
substructure of elements. This also implies the universal alternative choice
of gravity being either attractive or repulsive, which explains (up to
sign!)\ why the gravity is universally attractive.

3) The above conclusion does not exclude the possibility to interpret
black holes as gravitational sinks because they are \textit{singularity
points}. In the usual non-singular physical regions, however, physical 
detection of gravitational waves is made impossible by the absence of
gravitational sinks, which concludes the absence of gravitational waves 
at the experimentally observable levels.

4) Then, the notion of \textquotedblleft gravitons\textquotedblright\ and
\textquotedblleft quantization of gravity\textquotedblright\ are become
physically meaningless, in view of the essential roles played by such duality
relations as between wave and particle natures in the context of quantum theory.

As for the detailed accounts of the above remarks, please see my joint paper
\cite{OS} with Dr. H. Saigo, who has reformulated my idea of the absence
of gravitational sinks into that of gravitational \textquotedblleft black
bodies\textquotedblright\ to absorb gravitational field, according to which it
becomes evident that \textquotedblleft gravitational wave\textquotedblright%
\ cannot be used in the double slit experiment (because of the absence of
\textquotedblleft slits\textquotedblright\ for it), and hence, that it cannot
exhibit the interference effects. This implies the absence of wave characters
in the gravitational field.

\vskip10pt I would like to express my sincere thanks to Prof. Khrennikov for
his kind invitation to this exciting conference and for his kind hospitality
extended to me during it. I have also benefited very much from discussions
with him in Torun, Poland, after it.

\end{document}